\documentclass[aps,prb,twocolumn, showpacs]{revtex4}

\usepackage{epsfig}
\usepackage{graphicx}

\def\be{\begin{equation}}
\def\ee{\end{equation}}
\def\bea{\begin{eqnarray}}
\def\eea{\end{eqnarray}}

\begin{document}
\newcount\timehh  \newcount\timemm
\timehh=\time \divide\timehh by 60
\timemm=\time
\count255=\timehh\multiply\count255 by -60 \advance\timemm by \count255

\title{Spin currents in thermodynamic equilibrium?}
\author{Emmanuel I. Rashba\cite{Rashba*} }
\affiliation{
Department of Physics, SUNY at Buffalo, Buffalo, New York 14260, USA\\
and Department of Physics, MIT, Cambridge, Massachusetts 02139, USA}
\date{\today}

\begin{abstract}
The standard definition of a spin current, applied to the conductors lacking inversion symmetry, results in nonzero spin currents. I demonstrate that the spin currents do not vanish even in thermodynamic equilibrium, in the absence of external fields. These currents are dissipationless and are not associated with real spin transport. The result should be taken as a warning indicating problems inherent in the theory of transport spin currents 
driven by external fields. 

\end{abstract}

\maketitle

Generating spin currents (SCs) is one of the central goals of spintronics,\cite{review} and various mechanisms for electrical and optical injection of SCs have been proposed. Recently, the interest in SCs has been strengthened 
by independent reports of dissipationless SCs in two different systems:
holes in the valence band of a diamond type crystal described by 
a Luttinger Hamiltonian,\cite{Mur1} and electrons in a 2D system with 
a structure inversion asymmetry (SIA) described by a Rashba Hamiltonian.\cite{Sino} These systems differ in symmetry because the Luttinger Hamiltonian possesses inversion symmetry while the Rashba Hamiltonian lacks it. In both cases, SCs are driven by an external electric field \mbox{\boldmath$E$}. These surprising results have drawn a lot of attention, causing active interest and immediate response. They were followed by several papers of different researchers\cite{Schl,Hu,Shen} and also by the more recent papers coming from the same groups.\cite{Mur2,Cul,Sini} The mathematical formalism in some of these papers, particularly in Refs.~\onlinecite{Mur1}, \onlinecite{Hu} and \onlinecite{Mur2}, is rather involved. Meanwhile, the somewhat miraculous nature of the dissipationless SCs calls for a better understanding of their mechanism, including the properties of the background that supports the SCs linear in \mbox{\boldmath$E$}. From the standpoint of spintronic applications, it is important to understand whether these are {\it transport} currents, i.e., whether they can be employed for transporting spins, accumulating them at specific locations, and for injecting spins.

In this article I am trying to contribute to this basic physical understanding. I consider non-centrosymmetric 2D systems in thermodynamic equilibrium and show that using the standard definition of a SC results in non-vanishing SC expectation value. This result is not entirely surprising 
from a general symmetry viewpoint. Indeed, the reversal of the momenta (or the velocities) only, without reversing the angular momenta (spins), cannot be reconciled with the time-inversion symmetry for non-centrosymmetric systems.\cite{LP} Of course, such {\it background} currents, which are present in the ground state in the absence of in-plane external fields,
are non-transport currents. 

The standard Hamiltonian with a Rashba term is
\begin{equation}
H_R=\hbar^2k^2/2m+\alpha_R(\mbox{\boldmath$\sigma$}\times\mbox{\boldmath$k$})\cdot\hat{\bf z},
\label{eq1}
\end{equation}
with \mbox{\boldmath$\sigma$} the Pauli matrices, $\mbox{\boldmath$k$}=(k_x, k_y)$ the 2D momentum, and $\hat{\bf z}$ a unit vector perpendicular to the confinement plane. The eigenvalues of the Hamiltonian $H_R$ are 
$E_\lambda(k)=\hbar^2k^2/2m+\lambda\vert\alpha_R\vert k$, where $\lambda=\pm 1$ correspond to the upper and lower branches of the spectrum, respectively. The eigenfunctions are
\begin{equation}
\psi_\lambda (\mbox{\boldmath$k$})= {1\over \sqrt{2}}\left(
\begin{array}{c}
1\\-i\lambda\alpha_R (k_x+ik_y)/\vert\alpha_R\vert k)
\end{array}\right).
\label{eq2}
\end{equation}
The operator of the velocity is  
\begin{equation}
\mbox{\boldmath$v$}=\hbar^{-1}\partial H_R/\partial\mbox{\boldmath$k$}=\hbar\mbox{\boldmath$k$}/m+\alpha_R(\hat{\bf z}\times\mbox{\boldmath$\sigma$}),
\label{eq3}
\end{equation}
and the mean values of the Pauli matrices in the eigenstates $\psi_\lambda (\mbox{\boldmath$k$})$ are
\begin{equation}
\langle\mbox{\boldmath$\sigma$}\rangle_{\lambda\mbox{\boldmath$k$}}\equiv
\langle\psi_\lambda (\mbox{\boldmath$k$})\vert\mbox{\boldmath$\sigma$}\vert\psi_\lambda (\mbox{\boldmath$k$})\rangle=
\frac{\lambda\alpha_R}{\vert\alpha_R\vert k}(\mbox{\boldmath$k$}\times\hat{\bf z}).
\label{eq4}
\end{equation}
These equations allow one to evaluate the SC components defined as products of the electron velocity and spin components.

For calculating SCs in a given state $(\lambda, \mbox{\boldmath$k$})$, it is convenient to evaluate the Hermitian dot and cross products of \mbox{\boldmath$\sigma$} and \mbox{\boldmath$v$} by using Eqs.~(\ref{eq3}) and (\ref{eq4}):
\[{1\over 2}[(\mbox{\boldmath$\sigma$}\cdot\mbox{\boldmath$v$})+(\mbox{\boldmath$v$}\cdot\mbox{\boldmath$\sigma$})]={\hbar\over m}(\mbox{\boldmath$\sigma$}\cdot\mbox{\boldmath$k$}),
~\langle(\mbox{\boldmath$\sigma$}\cdot\mbox{\boldmath$k$})\rangle_{\lambda\mbox{\boldmath$k$}}=0,\]
and
\begin{equation}
{1\over 2}\langle[(\mbox{\boldmath$\sigma$}\times\mbox{\boldmath$v$})-(\mbox{\boldmath$v$}\times\mbox{\boldmath$\sigma$})]\rangle_{\lambda\mbox{\boldmath$k$}}
={\alpha_R\over\hbar}\left(2+\lambda{{\hbar^2k}\over {m\vert\alpha_R\vert}}\right).
\label{eq6}
\end{equation}
The dot-product vanishes because the spin is polarized perpendicularly to \mbox{\boldmath$k$} in the eigenstates $\psi_\lambda (\mbox{\boldmath$k$})$. The Kramers conjugate states, $\psi_\lambda (\mbox{\boldmath$k$})$ and $\psi_\lambda (-\mbox{\boldmath$k$})$, belong to the same branch of the spectrum. As a result, SCs of Eq.~(\ref{eq6}) are even with respect to the change $\mbox{\boldmath$k$}\rightarrow -\mbox{\boldmath$k$}$ within each branch, while their values for different branches, $\lambda=\pm 1$, are not mutually related. Therefore, there is no compensation of the SCs from different branches, and net macroscopic SCs can arise.

Electron spin is not conserved in the presence of a spin-orbit interaction because of precession in a momentum-dependent effective magnetic field. Therefore, the usual procedure for deriving currents from the continuity equations is not applicable to SCs  (at least in its simplest form), and I use in what follows the standard and physically appealing definition of the SC tensor ${\cal J}_{ij}$
\begin{equation}
{\cal J}_{ij}={1\over 2}\sum_\lambda\int{{d^2k}\over(2\pi)^2}\langle\sigma_iv_j+v_j\sigma_i\rangle_{\lambda\mbox{\boldmath$k$}}.
\label{eq7}
\end{equation}
Here $i,j=x,y$, with $i$ indicating the spin component and $j$ the transport direction. For $T=0$, the integration should be performed over $k\leq K_{\pm}$, where $K_{\pm}$ are the Fermi momenta for both spectral branches. For the Hamiltonian $H_R$ the tensor ${\cal J}_{ij}$ is antisymmetric
\begin{equation}
{\cal J}_{xx}={\cal J}_{yy}=0,~~{\cal J}_{xy}=-{\cal J}_{yx}\equiv{\cal J}_R.
\label{eq8}
\end{equation}
This tensor is invariant under the operations of the symmetry group $\mbox{\boldmath$C$}_{\infty v}$ of the Hamiltonian $H_R$.\cite{1D}

When the electrochemical potential $\mu$ is positive, $\mu>0$, both spectral 
branches are populated and
\begin{equation}
{\cal J}_R(\mu)=m^2\alpha_R^3/3\pi\hbar^5.
\label{eq9}
\end{equation}
Therefore, the SC is odd in the coupling constant $\alpha_R$, it is of the third order in $\alpha_R$, and does not depend on $\mu$. For comparison, the spin-orbit energy found by averaging the second term of $H_R$ is even in $\alpha_R$ ($\mu\geq 0$):
\begin{equation}
E_{\rm so}(\mu)=-(4/3\pi)(m\alpha_R/\hbar^2)^2(m\alpha_R^2/\hbar^2+3\mu/2).
\label{eq10}
\end{equation}

For small electron concentrations, when $\mu<0$ and electrons populate only the lower branch of the spectrum,
\begin{equation}
{\cal J}_R(\mu)={{m\alpha_R}\over{3\pi\hbar^3}}\left({{m\alpha_R^2}\over{\hbar^2}}-\mu\right)\sqrt{1+2{{\mu\hbar^2}\over{m\alpha_R^2}}}.
\label{eq11}
\end{equation}
Equations (\ref{eq9}) and (\ref{eq11}) match smoothly at $\mu=0$; a discontinuity exists only in the second derivative. Near the bottom of the spectrum, at $\mu=E_{\rm min}=-m\alpha_R^2/2\hbar^2$, ${\cal J}_{xy}(\mu)$ shows a square-root singularity. The nonanalytical behavior at these points emerges because of the spectrum singularities at $E=0$ and $E=E_{\rm min}$ and is also known for different phenomena.\cite{BR}

The SCs of Eqs.~(\ref{eq9}) and (\ref{eq11}) were found under the conditions of thermodynamic equilibrium. Therefore, they do not describe any real transport of electron spins and cannot result in spin injection or 
accumulation. Calculating transport currents would require a modification of Eq.~(\ref{eq7}). Nevertheless, Eq.~(\ref{eq8}) provides some insight into the effect of spin-orbit interaction on an equilibrium electron system. A real (i.e., invariant under time inversion) antisymmetric pseudo-tensor ${\cal J}_{ij}$ is equivalent to a real vector $\mbox{\boldmath$P$}\parallel\hat{\bf z}$. It has the symmetry of the normal electric field $\mbox{\boldmath$E$}_\perp$ producing the SIA and can be related to the spin-orbit contribution to the response of 2D electrons to this field. Interestingly, electric fields generated by SCs in magnetic insulators were discussed by Meier and Loss.\cite{ML} (Despite a similarity, both the origin and the scale of the effect are very different from the above.)

Calculating SCs driven by an in-plane field \mbox{\boldmath$E$} is beyond
the scope of the present work. Nevertheless, we point out at a resemblance between the background SCs ${\cal J}_{xy}=-{\cal J}_{yx}$ of Eq.~(\ref{eq8}) produced by the field $\mbox{\boldmath$E$}_\perp$ and the dissipationless SCs of Refs.~\onlinecite{Mur1} and \onlinecite{Sino}. Applying a driving field $\mbox{\boldmath$E$}$ to a diamond type crystal violates its inversion symmetry and, therefore, can produce background currents. To obtain
transport SCs, these background currents should be eliminated. 

The symmetry of the tensor ${\cal J}_{ij}$ depends on the specific choice of the spin-orbit interaction. When it originates from the bulk inversion asymmetry (BIA), the Hamiltonian includes a Dresselhaus spin-orbit term. In the principal cubic axes it reads
\begin{equation}
H_D=\hbar^2k^2/2m+\alpha_D(\sigma_xk_x-\sigma_yk_y).
\label{eq12}
\end{equation}
As distinct from $H_R$, $H_D$ possesses only $\mbox{\boldmath$C$}_{2v}$ symmetry. As a result, the ordering of electron spins with respect to \mbox{\boldmath$k$} is quite different, and
\begin{equation}
{\cal J}_{xx}=-{\cal J}_{yy}={\cal J}_D,~~{\cal J}_{xy}={\cal J}_{yx}=0,
\label{eq13}
\end{equation}
where ${\cal J}_D(\mu)$ can be found from ${\cal J}_R(\mu)$ of Eqs.~(\ref{eq9}) and (\ref{eq11}) by the substitution $\alpha_R\rightarrow\alpha_D$. Therefore, Eq.~(\ref{eq7}) results in equilibrium SCs also for 
BIA systems.

In conclusion, the standard procedure for calculating spin-currents, when applied to non-centrosymmetric crystals, results in non-vanishing
currents even under the conditions of thermodynamic equilibrium. 
For calculating transport spin-currents, a procedure for eliminating 
the background currents should be devised.

The financial support from DARPA/SPINS by the ONR Grant N000140010819 is gratefully acknowledged. I am grateful to L. S. Levitov for a critical discussion. A. G. Mal'shukov kindly informed me that he arrived at
some of the above results independently.

\end{document}